\documentclass[preprint,12pt]{elsarticle}%
\usepackage{amssymb}
\usepackage{amsfonts}
\usepackage{amsmath}
\usepackage{graphicx}%
\providecommand{\U}[1]{\protect\rule{.1in}{.1in}}
\journal{journal}

\begin{document}
\bigskip\bigskip%
\begin{frontmatter}%


%

\title
{Ultrastrong adhesion in the contact with thin elastic layers on a rigid foundation}%

%

\author{M. Ciavarella, A.Papangelo}%
%

\address
{Politecnico di BARI. Center of Excellence in Computational Mechanics. Viale Gentile 182, 70126 Bari. Mciava@poliba.it}%
%

\begin{abstract}%

In the present short note, we generalize simple approximate
Johnson-Jaffar-Barber solutions for the indentation by a rigid punch of a thin
elastic layer on a rigid foundation to the case of adhesion. This could be an
interesting geometry for an adhesive system, a limit case of the more general
class of layered systems, or FGMs (Functionally Graded\ Materials). We show
that ultrastrong adhesion (up to theoretical strength) can be reached both in
line contact or in axisymmetric contact for thin layers (typically of
nanoscale size), which suggests a new possible strategy for "optimal
adhesion". In particular, in line contact adhesion enhancement occurs as an
increase of the actual pull-off force, while in axisymmetric case the latter
is apparently very close to the classical JKR case. However, it appears in
closer examination that also for axisymmetric case, the enhancement occurs by
reducing the size of contact needed to sustain the pull-off force. These
effects are further enhanced by Poisson's ratio effects in the case of nearly
incompressible layer.%

\end{abstract}%
%

%

\end{frontmatter}%



\section{\bigskip Introduction}

Inspired by nature, we are studying and possibly imitating many ways to
optimize adhesion devices, one way being by reducing length scales involved in
the geometry (Gao \& Yao, 2004). A significant amount of study has been
devoted to the case of halfspace geometry, for which the optimal shape for
maximum pulloff force is found to be concave, although it is not "robust" to
surface geometry errors (Yao and Gao, 2006). Enhancement of adhesion due to
surface geometries is also known in mushroom-shaped fibrils (Peng and Cheng,
2012), rodlike particles (Sundaram and Chandrasekar, 2011), or moving to
functionally graded materials (FGMs) which are increasingly used in
engineering, and have been also used in nature as a result of evolution
(Suresh, 2001, Sherge \& Gorb, 2001). Indeed, few authors have explored the
behaviour of attachments using FGMs (Chen et al., 2009a, 2009b, Jin et al.,
2013), finding interesting results and possible avenues to design "optimal"
adhesive systems.

\bigskip However, curiously a much simpler geometry (which is in a sense a
limit case of FGM) is that of adhesion with a layer on a rigid foundation. 

In his well known book, Johnson (1985) suggested an elementary formulation to
obtain asymptotic results for the contact pressure between a frictionless
rigid indenter and a thin elastic layer supported by a rigid foundation.
Jaffar (1989) later on used the same technique for the axisymmetric case, and
finally Barber (1990) generalized it to the arbitrary, three-dimensional
problem for the thin elastic layer. 

A typical assumption made is that of the JKR model (Johnson et al., 1971)
which corresponds to very short range adhesion where adhesive forces are all
within the contact area. Solving the JKR problem is simple generalizing the
original JKR energetic derivation assuming calculation of the strain energy in
adhesiveless contact, and unloading at constant contact area (see (Ciavarella,
2017)). The underlying assumption of (Ciavarella, 2017) is that the contact
area distributions are the same as under adhesiveless conditions (for an
appropriately increased normal load). There are no approximations involved if
the geometry is that of a single line or axisymmetric contact, as the solution
is exact within the JKR assumption of infinitely short adhesion range, and
states that the indentation under adhesive conditions for a given surface
energy $w$ is%
\begin{equation}
\delta=\delta_{1}-\sqrt{2wA^{\prime}/P_{1}^{\prime\prime}}\label{ciavarella1}%
\end{equation}
where $\delta_{1}$ is the adhesiveless indentation, $A^{\prime}$ is the first
derivative of contact area and $P_{1}^{\prime\prime}$ the second derivative of
the adhesiveless load with respect to $\delta_{1}$. Then, the adhesive load
is
\begin{equation}
P=P_{1}-P_{1}^{\prime}\sqrt{2wA^{\prime}/P_{1}^{\prime\prime}}%
\label{ciavarella2}%
\end{equation}

Hence, the asymptotic solutions for the adhesive thin layer problems are found
quite simply from the solutions of Johnson (1985), Jaffar (1989), Barber
(1990). We shall then discuss implications, particularly as some results will
be quite surprising, and suggest potential strategies for "optimal" adhesion.

\bigskip

\section{The solution}

We follow Barber (1990) in denoting the contact surface of the layer by $z=0$
and choose a two-dimensional Cartesian coordinate system $x_{1},x_{2}$ in the
plane of the layer. The original Johnson's approximation is to assume that
plane sections within the layer remain plane, so that the in-plane
displacements of the layer with components $u_{1}$, $u_{2}$ is independent of
$z$. Barber (1990) shows that in the case the layer is on \textit{a
frictionless foundation} we recover the contact pressure %

\begin{equation}
p\left(  x_{1},x_{2}\right)  =\frac{2\mu}{\left(  1-\nu\right)  }w\left(
x_{1},x_{2}\right)
\end{equation}
where the indentation $w$ is the local interpenetration between the indenter
and the layer if it did not deform, $\mu$ is shear modulus and $\nu$ Poisson's
ratio. We could transform the adhesionless solution into an adhesive one with
an approximation following (Ciavarella, 2017) and the approximation will not
be the same as that used for the halfspace by Johnson \& Greenwood (2005)
which shows the ellipticity of the contact area with adhesion differs from
that without.

However, there are no approximations involved in applying (Ciavarella, 2017)
and hence (\ref{ciavarella1}, \ref{ciavarella2}) in the case of a single line
contact, or axisymmetric contact, which we shall consider next. We are already
solving the problem with the "asymptotic" assumption for the layer thickness,
and prefer in fact not to make further approximations.

\subsection{Line contact}

\subsubsection{Frictionless foundation}

As in the original Johnson (1985) derivation, we consider (see Fig.1) a layer
indented by a frictionless rigid cylinder of radius $R$, and assume the
thickness of the layer $b$ is small compared with the half-width of the
contact size $a,$ i.e. $b<<a,$ which is required for the assumptions that
plane sections remain plane after compression.

\begin{center}
\bigskip%
\begin{tabular}
[c]{ll}%
{\includegraphics[
height=3.2312in,
width=5.056in
]%
{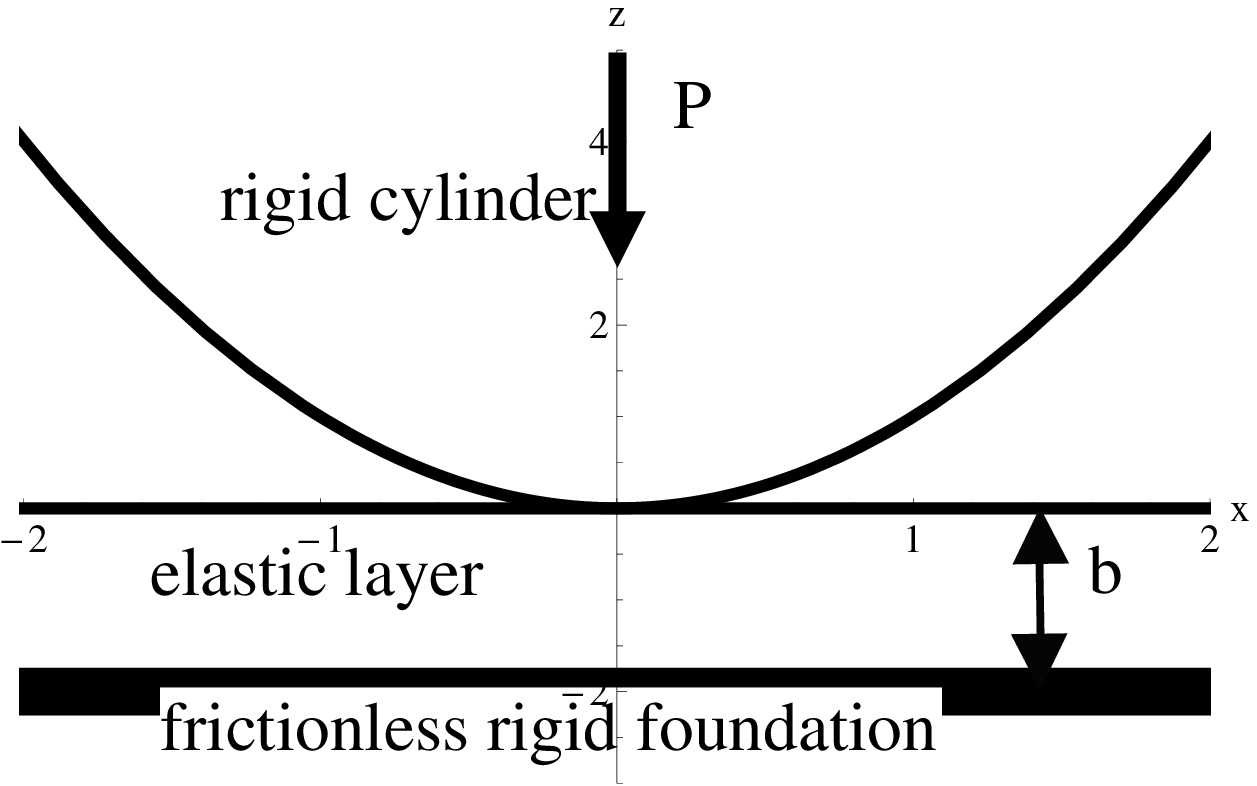}%
}
&
\end{tabular}

Fig 1. The geometry for a rigid cylinder indenting an layer on a frictionless
rigid foundation
\end{center}

The adhesiveless solution gives for indentation $\delta_{1}$ and load $P_{1}$%
\begin{align}
\delta_{1}  & =a^{2}/2R\\
P_{1}  & =\frac{2}{3}\frac{E^{\ast}L}{Rb}a^{3}=\frac{2^{5/2}}{3}\frac{E^{\ast
}LR^{1/2}}{b}\delta_{1}^{3/2}%
\end{align}
being $E^{\ast}$ the plane strain elastic modulus, $L$ the contact length. Now
the adhesive solution is obtained considering $A=2aL\ $and with obvious
algebra using (\ref{ciavarella2})
\begin{equation}
P=\frac{4}{3}\frac{E^{\ast}L\sqrt{2R\delta_{1}}}{b}\left(  \delta_{1}%
-3\sqrt{\frac{b}{2E^{\ast}}w}\right)  \label{load-layer}%
\end{equation}
in terms of the adhesionless indentation $\delta_{1}$. To find the minimum
load (pull-off), the condition $P^{\prime}=0$ gives%
\begin{equation}
\delta_{1,PO}=\sqrt{\frac{b}{2E^{\ast}}w}\quad;\quad a_{PO}=\sqrt{2R}\left(
\frac{b}{2E^{\ast}}w\right)  ^{1/4}%
\end{equation}
(where notice that we have to assume $a_{PO}>>b$ to be consistent with the
thin layer assumption), and hence substituting
\begin{equation}
P_{PO}=-\frac{8}{3}\frac{E^{\ast}LR^{1/2}}{(2b)^{1/4}}\left(  \frac{w}%
{E^{\ast}}\right)  ^{3/4}%
\end{equation}
whereas the average stress in the contact at pull-off is
\begin{equation}
\sigma_{PO}=\frac{P_{PO}}{2A_{PO}}=-\frac{2}{3}\sqrt{2}\left(  \frac{E^{\ast
}w}{b}\right)  ^{1/2}=-\frac{2}{3}\sqrt{2}\frac{K_{Ic}}{\sqrt{b}}%
\end{equation}
where $K_{Ic}$ is toughness of the contact. Hence, notice that the JKR
solution simply gives the Griffith condition imposed by a Stress Intensity
Factor which scales only with the size the layer $b$ and not any other length
scale (like the radius of the punch). The interesting result that as
$b\rightarrow0$ the limit of the force also goes to $\infty$. Since this will
be bounded by theoretical strength, the situation is analogous to the well
known case of a fibrillar structure in contact with a rigid halfspace, like
that discussed for Gecko and many insects who have adopted nanoscale fibrillar
structures on their feet as adhesion devices (Gao \& Yao, 2004). In our case,
to have a design insensitive to small variations in the tip shape, we would
simply need to go down in the scale of the layer thickness. In fact, by
equating $\sigma_{PO}$ to \ theoretical strength, we obtain the order of
magnitude of the "critical" thickness below which we expect theoretical
strength%
\begin{equation}
b_{cr}=\frac{8}{9}\frac{E^{\ast}w}{\sigma_{th}^{2}}%
\end{equation}
Taking $w=10mJ/m^{2},\sigma_{th}=20MPa$ and $E^{\ast}=1GPa$, like done in (Gao
\& Yao, 2004), we estimate $b_{cr}=\frac{8}{9}\frac{10^{9}10^{-2}}{\left(
20\times10^{6}\right)  ^{2}}=22nm$ , which is quite similar to the estimate
(of a different geometry) of 64$nm$ robust design diameter of the fiber of the
fibrillar structure. Hence, with this size of layer of nanoscopic scale, we
would be able to devise a quite strong attachment for any indenter. 

In the halfplane limit case, from Barquins (1988), Chaudhury et al. (1996) we
have for the cylinder%

\begin{align}
P_{PO,HP}  & =-3L\left(  \frac{\pi E^{\ast}w^{2}R}{16}\right)  ^{1/3}%
\quad;\quad a_{PO,HP}=\left(  \frac{2wR^{2}}{\pi E^{\ast}}\right)  ^{1/3}\\
\sigma_{PO,HP}  & =\frac{P_{PO,HP}}{2A_{PO,HP}}=\frac{-3}{2}\left(  \frac
{\pi^{2}E^{\ast2}w}{32R}\right)  ^{1/3}%
\end{align}
which does include some dependence on elastic modulus which is not present in
the axisymmetric halfspace problem of JKR model (Johnson et al., 1971), but it
seems to be quite different in terms of power law dependence from the
"layered" case. Indeed, take the ratio%
\begin{equation}
\frac{P_{PO}}{P_{PO,HP}}=\frac{8}{9}\frac{16^{1/3}}{\pi^{1/3}2^{1/4}}%
\frac{R^{1/6}}{b^{1/4}}\left(  \frac{w}{E^{\ast}}\right)  ^{1/12}%
\end{equation}
which shows how there are really different power law dependences in the layer limit.

The full curve $P-\delta$ is then obtained using (\ref{ciavarella1})
\begin{equation}
\delta=a^{2}/2R-\sqrt{2w\frac{b}{E^{\ast}}}%
\end{equation}
so extracting the equation for the contact area, using $\delta_{1}=a^{2}/2R$,
and then substituting back in the solution (\ref{load-layer}), we get
\[
\widehat{P}=\frac{4}{3}\sqrt{2}\left(  \widehat{\delta}+\sqrt{2}\right)
^{1/2}\left(  \widehat{\delta}+\sqrt{2}-\frac{3}{\sqrt{2}}\right)
\]
where we have defined dimensionless quantities%
\[
\widehat{\delta}=\frac{\delta}{\sqrt{w\frac{b\ }{E^{\ast}\ }}}\quad
;\quad\widehat{P}=\frac{P}{\frac{E^{\ast}LR^{1/2}}{b^{1/4}}\left(  \frac
{w}{E^{\ast}}\right)  ^{3/4}}%
\]
so that $\widehat{P}_{PO}=-\frac{8}{3\times2^{1/4}}=2.\,\allowbreak242\,4$ and
$\widehat{\delta}_{PO}=-\frac{\sqrt{2}}{2}=0.707\,$. 

Following Fig.2, the solution is plotted in dimensionless terms, and typical
loading paths are indicated by arrows: starting from remote locations, one
finds contact only when there is contact of the undeformed surfaces (JKR makes
it not possible to model long range adhesion) and hence until point O (the
origin of the coordinate system) is reached. Then under force control, one
would obtain a jump to point B where force remains zero but one finds an
effective indentation $\widehat{\delta}_{B}$. From this point on, one could
load in compression and go up in the figure, or start downloading. Unloading
then ends at pull-off in point $\widehat{P}_{PO}$. Alternatively, if we were
under displacement control, at the point of first contact we would build up
adhesive force and jump to point A, with an effective tensile load but then
unloading would proceed until the adhesive force is reduced back to zero in
point C. Hence, there is no pull-off under displacement control, contrary to
the classical JKR case.

\begin{center}
\bigskip%
\begin{tabular}
[c]{ll}%
{\includegraphics[
height=3.2461in,
width=5.056in
]%
{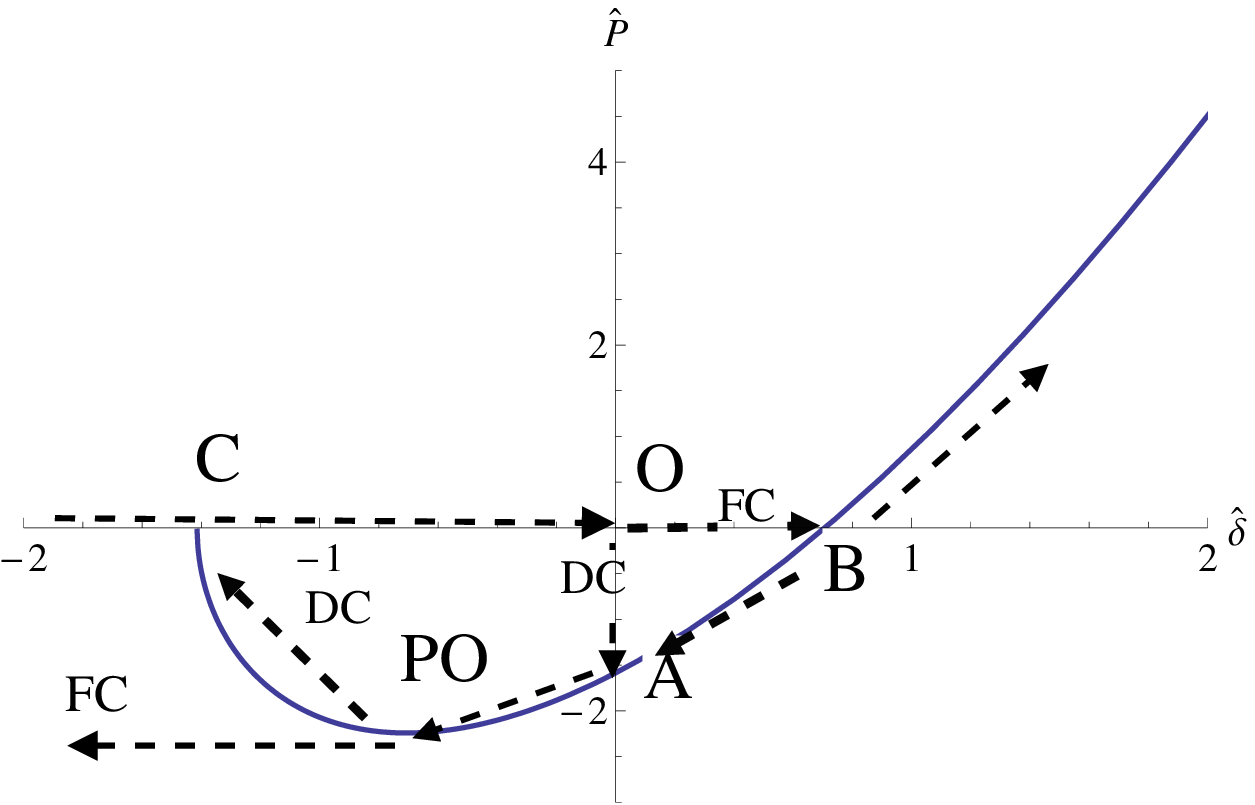}%
}
&
\end{tabular}

Fig 2. Load vs indentation curve for a rigid cylinder indenting an layer on a
frictionless rigid foundation. Possible loading path under Force Control (LC)
and Displacement Control (DC) 
\end{center}

\bigskip

\subsubsection{Bonded layer}

For bonded layer, a similar procedure finds
\begin{align}
P_{PO}  & =-\frac{8}{3}\frac{E^{\ast}LR^{1/2}}{(2b)^{1/4}}\frac{\left(
1-\nu\right)  ^{1/2}}{2^{1/4}\left(  1-2\nu\right)  ^{1/4}}\left(  \frac
{w}{E^{\ast}}\right)  ^{3/4}\\
a_{PO}  & =\sqrt{2R}\left(  \frac{\left(  1-2\nu\right)  }{\left(
1-\nu\right)  ^{2}}\frac{b}{E^{\ast}}w\right)  ^{1/4}%
\end{align}
and therefore for the bonded layer the Poisson's effect appears, which only
changes a prefactor in the result for the frictionless foundation --- but
notice this prefactor makes the load diverge towards the incompressible limit
$\nu=0.5$. \bigskip Hence, in this case the average stress in the contact at
pull-off is
\begin{equation}
\sigma_{PO}=\frac{P_{PO}}{2A_{PO}}=-\frac{4}{3}\frac{\left(  E^{\ast}w\right)
^{1/2}}{b^{1/2}}\frac{1-\nu}{2\left(  1-2\nu\right)  ^{1/2}}%
\end{equation}
and hence here by equating $\sigma_{PO}$ to \ theoretical strength, we obtain%
\begin{equation}
b_{cr}=\left[  \frac{1}{2}\frac{\left(  1-\nu\right)  ^{2}}{1-2\nu}\right]
\frac{8}{9}\frac{E^{\ast}w}{\sigma_{th}^{2}}=\left[  \frac{1}{2}\frac{\left(
1-\nu\right)  ^{2}}{1-2\nu}\right]  b_{cr,frictionless}%
\end{equation}
and therefore this time the critical layer thickness becomes dependent on
Poisson's ratio, rendering the layer adhesive \textit{much more effective}.

\subsubsection{\bigskip Incompressible bonded layer}

The results of the previous paragraph hold until the layer is nearly
incompressible, in which case a similar procedure yields
\begin{equation}
P_{PO}=-\frac{8}{5}L\frac{\left(  3Rw\right)  ^{2/3}}{\left(  2b\right)
^{1/2}}w^{1/6}%
\end{equation}
and $\delta_{1,PO}=b\left(  \frac{w}{3E^{\ast}R}\right)  ^{1/3}$ while
$a_{PO}=\sqrt{6Rb\left(  \frac{w}{3E^{\ast}R}\right)  ^{1/3}}$, which is
therefore rather different from the frictionless counterpart. Hence, in this
case the average stress in the contact at pull-off is
\[
\sigma_{PO}=\frac{P_{PO}}{2A_{PO}}=-\frac{4}{5}\frac{3^{5/6}}{\left(
12\right)  ^{1/2}}\frac{w^{2/3}E^{\ast1/6}}{b}R^{1/3}%
\]
and we return to see effects of the radius of the indenter (i.e. qualitative
effects on the geometry) like in the halfplane problem. 

\section{Axysimmetric problems}

Jaffar (1989) has studied axisymmetric contact problems involving bonded and
unbounded layers indented by a frictionless rigid body. For the case of
unbounded frictionless foundation, the total load is given by
\begin{equation}
P_{1}=\frac{\pi}{4}E^{\ast}\frac{a^{4}}{Rb}%
\end{equation}

For the value of the penetration depth $\delta_{1}$ when $a/b>>1$, no
asymptotic formula exists as for line contacts; but it can be obtained on the
interpretation that "it represents zero change in volume of the material under
the indenter"
\begin{equation}
\delta_{1}=\frac{a^{2}}{4R}%
\end{equation}
Hence we have for the repulsive solution $P_{1}=4\pi E^{\ast}\frac{R}{b}%
\delta_{1}^{2}$ and therefore repeating the standard procedure, the adhesive
solution is obtained as
\begin{equation}
P=P_{1}-P_{1}^{\prime}\sqrt{2wA^{\prime}/P_{1}^{\prime\prime}}=4\pi E^{\ast
}\frac{R}{b}\delta_{1}\left(  \delta_{1}-2\sqrt{\frac{w}{E^{\ast}}b}\right)
\label{load-layer3D}%
\end{equation}
Finding the minimum,
\begin{equation}
\delta_{1,PO}=\sqrt{\frac{w}{E^{\ast}}b}\quad;\quad a_{PO}=\sqrt{4R}\left(
\frac{w}{E^{\ast}}b\right)  ^{1/4}%
\end{equation}
(where notice that we have to assume $a_{PO}>>b$ to be consistent with the
thin layer assumption). Hence, substituting $\delta_{1,PO}$
\begin{equation}
P_{PO}=-4\pi Rw
\end{equation}
and simply this is much larger than the JKR classical solution, but shows the
same independence on elastic modulus, and, surprisingly, also on thickness of
the layer (contrary to the line contact case). Notice however that
\begin{equation}
\sigma_{PO}=\frac{P_{PO}}{\pi a_{PO}^{2}}=\frac{\sqrt{wE^{\ast}}}{\sqrt{b}%
}=\frac{K_{Ic}}{\sqrt{b}}%
\end{equation}
and equalling $\sigma_{PO}=\sigma_{th}$ we get
\begin{equation}
b_{cr}=\frac{E^{\ast}w}{\sigma_{th}^{2}}%
\end{equation}
which is of the same order of $b_{cr,frictionless}$ in the line contact case.
Although the pull-off load seems close to the JKR value and does not depend on
layer thickness, this can be made still "optimal" as it is clear from a more
complete comparison with the JKR\ classical solution
\begin{equation}
P_{JKR}=\sqrt{8\pi E^{\ast}wa^{3}}-\frac{4}{3}E^{\ast}\frac{a^{3}}{R}%
\end{equation}
whose minimum (force control) is at $P_{JKR,PO}=-\frac{3}{2}\pi Rw$ and
$a_{JKR,PO}^{3}=\frac{9}{8}\pi\frac{R^{2}w}{E^{\ast}}$. Hence, the ratio of
contact areas at pull-off%
\begin{equation}
\frac{a_{PO}}{a_{JKR,PO}}=\frac{\sqrt{4}}{\sqrt[3]{\frac{9\pi}{8}}}%
\frac{b^{1/4}}{R^{1/6}\left(  \frac{w}{E^{\ast}}\right)  ^{1/12}}%
\end{equation}
and hence this shows for thin layers, the same pull-off load of the JKR
solution is reached with a much smaller contact area (tending to zero --- but
should not be smaller than thickness for not violating the thin-film
assumption): this therefore could serve to optimize the adhesive system. In
particular, for the "optimal" layer thickness, the ratio of contact areas at
pull-off%
\begin{equation}
\left(  \frac{a_{PO}}{a_{JKR,PO}}\right)  _{b=b_{cr}}=\frac{\sqrt{4}}%
{\sqrt[3]{\frac{9\pi}{8}}}\frac{\left(  \frac{E^{\ast}}{\sigma_{th}}\right)
^{1/2}\left(  \frac{w}{E^{\ast}R}\right)  ^{1/6}}{\left(  w/E^{\ast}\right)
^{1/12}}%
\end{equation}

Therefore taking $w=10mJ/m^{2},\sigma_{th}=20MPa$ and $E^{\ast}=1GPa$, again
as in the example of (Gao \& Yao, 2004), we estimate $\allowbreak b_{cr}%
=\frac{10^{9}10^{-2}}{\left(  20\cdot10^{6}\right)  ^{2}}=\allowbreak
2.5\times10^{-8}m$. However, we also have to check that this is feasible
within the thin layer assumption. However, the two conditions, $a_{PO}>>b$
(thin layer) and $\frac{a_{PO}}{a_{JKR,PO}}<<1$ (optimal adhesion) lead to the
same dependences
\begin{align*}
b  & <<2^{4/3}R^{2/3}\left(  \frac{w}{E^{\ast}}\right)  ^{1/3}\text{ (thin
layer)}\\
b  & <<\frac{1}{2^{4}}\left(  \frac{9\pi}{8}\right)  ^{4/3}R^{2/3}\left(
\frac{w}{E^{\ast}}\right)  ^{1/3}\text{ (optimal adhesion)}%
\end{align*}
of which the second dominates. Hence, indeed, having $b<b_{cr}$ leads to a
feasible regime of optimal adhesion.

Turning back to the complete solution, the indentation is obtained using
(\ref{ciavarella1})
\begin{equation}
\delta=\frac{a^{2}}{4R}-\sqrt{w\frac{b}{E^{\ast}}}%
\end{equation}
and so extracting the equation for the contact area, and $\frac{a^{2}}%
{4R}=\delta+\sqrt{w\frac{b}{E^{\ast}}}$, using $\delta_{1}=\frac{a^{2}}%
{4R}=\delta+\sqrt{w\frac{b}{E^{\ast}}}$ and substituting back in the solution
(\ref{load-layer3D}), we obtain the quite simple relationship between adhesive
load and indentation%
\begin{equation}
\widehat{P}_{axi}=\widehat{\delta}^{2}-1
\end{equation}
using the same dimensionless notation of line contact for $\widehat{\delta}$,
but obviously not that for $\widehat{P}$%
\begin{equation}
\widehat{\delta}=\frac{\delta}{\sqrt{w\frac{b\ }{E^{\ast}\ }}}\quad
;\quad\widehat{P}_{axi}=\frac{P}{4\pi Rw}%
\end{equation}
so that $\widehat{P}_{PO}=-1$ and $\widehat{\delta}_{PO}=0$. This is plotted
in Fig.3 with the remark that $\delta_{PO}=0$ and that there seems to be no
instability under displacement control on separation. 

\begin{center}
\bigskip%
\begin{tabular}
[c]{ll}%
{\includegraphics[
height=3.2461in,
width=5.056in
]%
{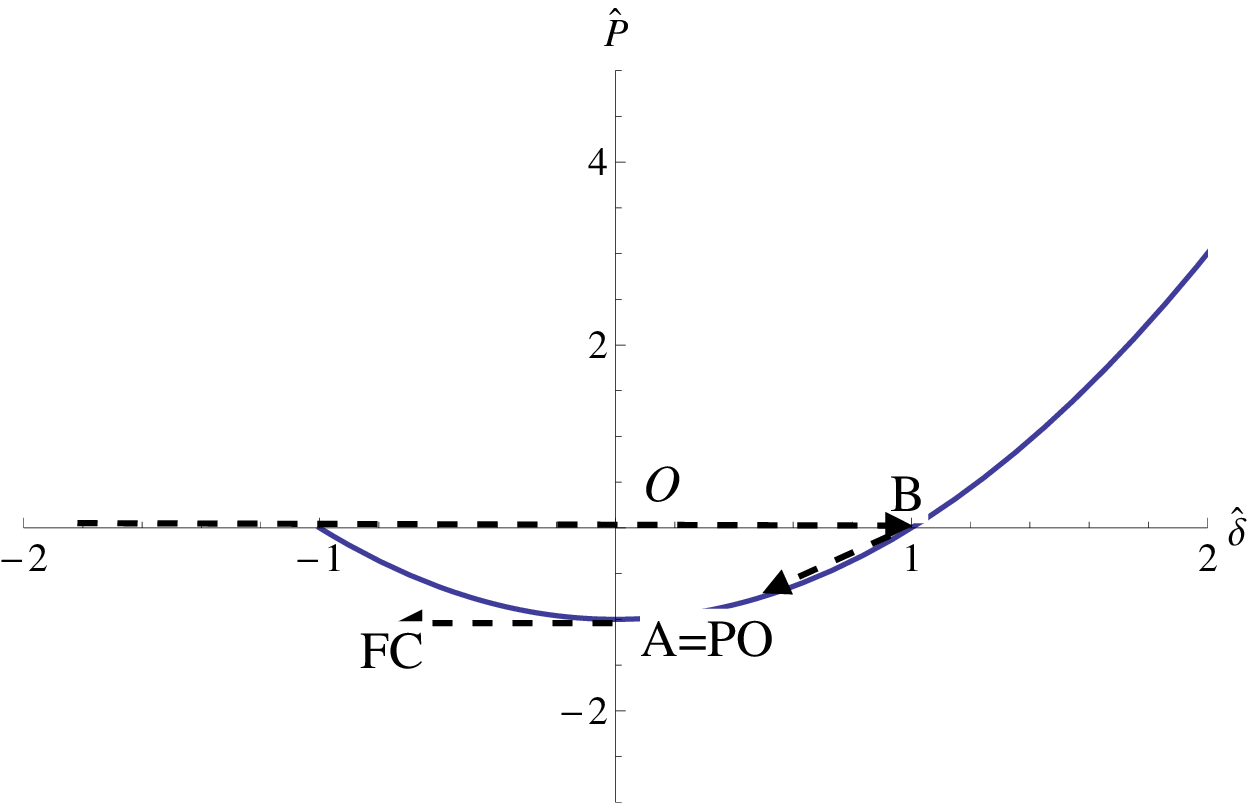}%
}
&
\end{tabular}

Fig.3. Load vs indentation curve for a rigid sphere indenting a layer on a
frictionless rigid foundation. Only force control path is indicated.
\end{center}

\bigskip

\subsection{Bonded layer}

For bonded layer the Poisson's effect appears. In fact, Jaffar (1989) shows
\begin{equation}
P_{1}=\frac{\pi}{4}E^{\ast}\frac{\left(  1-\nu\right)  ^{2}}{1-2\nu}%
\frac{a^{4}}{Rb}%
\end{equation}
whereas we continue to use $\delta_{1}=\frac{a^{2}}{4R}$. Therefore,
$P_{1}=4\pi E^{\ast}\frac{\left(  1-\nu\right)  ^{2}}{1-2\nu}\frac{R}{b}%
\delta_{1}^{2}$ and the adhesive solution is obtained as
\begin{equation}
P=P_{1}-P_{1}^{\prime}\sqrt{2wA^{\prime}/P_{1}^{\prime\prime}}=4\pi E^{\ast
}\frac{\left(  1-\nu\right)  ^{2}}{1-2\nu}\frac{R}{b}\left(  \delta_{1}%
^{2}-2\delta_{1}\sqrt{\frac{w}{\frac{\left(  1-\nu\right)  ^{2}}{1-2\nu
}E^{\ast}}b}\right)
\end{equation}
Finding the minimum,
\begin{equation}
\delta_{1,PO}=\sqrt{\frac{\left(  1-2\nu\right)  w}{\left(  1-\nu\right)
^{2}E^{\ast}}b}\quad;\quad a_{PO}=\sqrt{4R}\left(  \frac{\left(
1-2\nu\right)  w}{E^{\ast}\left(  1-\nu\right)  ^{2}}b\right)  ^{1/4}%
\end{equation}
and hence we obtain the same pull-off of the frictionless foundation,
\begin{equation}
P_{PO}=-4\pi Rw
\end{equation}
It is only the area of contact at pull-off which reduces and therefore the
contact is indeed more "efficient" than the frictionless case.

\subsubsection{Bonded incompressible layer}

The last case that needs to be considered is the bonded case with
incompressible layer, $\nu=0.5.$From Jaffar (1989) we have for the repulsive
solution
\[
P_{1}=\frac{\pi}{3\times2^{5}}E^{\ast}\frac{a^{6}}{Rb^{3}}%
\]
therefore
\begin{equation}
P=\frac{2}{3}\pi E^{\ast}\frac{R^{2}}{b^{3}}\delta_{1}^{3/2}\left(  \delta
_{1}^{3/2}-3\sqrt{\frac{2w}{RE^{\ast}}b^{3}}\right)
\end{equation}
and pull-off is found at%
\begin{equation}
\delta_{1,PO}=\left(  \frac{9b^{3}w}{2E^{\ast}R}\right)  ^{1/3}%
\end{equation}
and hence
\begin{equation}
P_{PO}=-3\pi Rw
\end{equation}
Notice that for the case of incompressible layer the pull-off is lower and the
indentation at pull-off depends on the radius of the sphere.

\section{Conclusions}

In this short comunication, we show that ultrastrong adhesion can be reached
both in line contact or in axisymmetric contact for contact of a Hertzian
indenter with ultrathin layers, suggesting a new possible strategy for
"optimal adhesion". There are some details which differ in plane contact vs
axysimmetrical contact: indeed, in line contact adhesion enhancement occurs as
an increase of the actual pull-off force, while in axisymmetric case the
latter is apparently very close to the classical JKR case. However, in both
cases the enhancement occurs because the dominant length scale for the stress
intensity factor at the contact edge is the layer thickness, and this induces
a reduction of the size of contact needed to sustain the pull-off force. These
effects are remarkably further enhanced by Poisson's ratio effects in the case
of nearly incompressible layer.

\section{References}

Barber, J. R. (1990). Contact problems for the thin elastic layer.
International journal of mechanical sciences, 32(2), 129-132.

Barquins, M. (1988). Adherence and rolling kinetics of a rigid cylinder in
contact with a natural rubber surface. The Journal of Adhesion, 26(1), 1-12.

Chaudhury, M. K., Weaver, T., Hui, C. Y., \& Kramer, E. J. (1996). Adhesive
contact of cylindrical lens and a flat sheet. Journal of Applied Physics,
80(1), 30-37.

Chen, S., Yan, C., and Soh, A., (2009a), \textquotedblleft Adhesive Behavior
of Two-Dimensional Power-Law Graded Materials,. Int. J. Solids Struct., 46,
pp. 3398--3404.

Chen, S., Yan, C., Zhang, P., and Gao, H., (2009b), \textquotedblleft
Mechanics of Adhesive Contact on a Power-Law Graded Elastic
Half-Space,\textquotedblright\ J. Mech. Phys. Solids, 57, pp.1437--1448.

Ciavarella, M. (2017). An approximate JKR solution for a general contact,
including rough contacts. arXiv preprint arXiv:1712.05844.

Fuller, K. N. G., \& Tabor, D. The effect of surface roughness on the adhesion
of elastic solids. Proc Roy Soc London A: 1975; 345:1642, 327-342

Gao, H., \& Yao, H. (2004). Shape insensitive optimal adhesion of nanoscale
fibrillar structures. Proceedings of the National Academy of Sciences of the
United States of America, 101(21), 7851-7856.

Jaffar, M. J. (1989) Asymptotic behaviour of thin elastic layers bonded and
unbonded to a rigid foundation. Int. J.Mech. Sci. 31, 229 .

Jin, F., Guo, X., \& Zhang, W. (2013). A unified treatment of axisymmetric
adhesive contact on a power-law graded elastic half-space. Journal of Applied
Mechanics, 80(6), 061024.

Johnson, K. L., Contact Mechanics, pp. 138-141. Cambridge University Press,
Cambridge (1985). 

Johnson, K. L., \& Greenwood, J. A. (2005). An approximate JKR theory for
elliptical contacts. Journal of Physics D: Applied Physics, 38(7), 1042.

Johnson, K. L., K. Kendall, and A. D. Roberts. (1971). Surface energy and the
contact of elastic solids. Proc Royal Soc London A: 324. 1558.

\bigskip Peng, Z., and Chen, S., 2012, \textquotedblleft The Effect of
Geometry on the Adhesive Behavior of Bio-Inspired Fibrils,\textquotedblright%
\ Soft Matter 8, pp. 9864--9869.

Yao, H., \& Gao, H. (2006). Optimal shapes for adhesive binding between two
elastic bodies. Journal of colloid and interface science, 298(2), 564-572.

\bigskip Sherge, M., and Gorb, S., 2001, Biological Micro- and Nano-Tribology-
Nature's Solutions, Springer, Berlin

Sundaram, N., and Chandrasekar, S., 2011, \textquotedblleft Shape and
Eccentricity Effects in Adhesive Contacts of Rodlike
Particles,\textquotedblright\ Langmuir 27, pp. 12405--12410

\bigskip Suresh, S., 2001, \textquotedblleft Graded Materials for Resistance
to Contact Deformation and Damage,\textquotedblright\ Science, 292, pp. 2447--2451.

\end{document}